\documentclass[epj]{svjour}
\usepackage{graphics}
\usepackage[utf8]{inputenc}
\usepackage{xcolor}
\usepackage{breakcites}
\usepackage{amsmath}
\usepackage{rotating}
\usepackage{dcolumn}
\usepackage{bm}
\usepackage{breakcites}

\begin{document}

\title{Effects of Strong Magnetic Fields on the Hadron-Quark Deconfinement Transition}

\author{Betânia C. T. Backes \and
Kauan D. Marquez \and
Débora P. Menezes}

\institute{Departamento de F\'isica, CFM - Universidade Federal de Santa Catarina; C.P. 476, CEP 88.040-900, Florian\'opolis, SC, Brasil }

\date{Received: date / Revised version: date}

\abstract{
The aim of the present work is to investigate the effects of strong magnetic fields on the hadron-quark phase transition point at zero temperature. To describe the hadronic phase, a relativistic mean field (RMF) model is used and to describe the quark phase a density dependent quark mass model (DDQM) is employed. As compared with the results obtained with non-magnetised matter, we observe a shift of the transition point towards higher pressures and, generally also towards higher chemical potentials. An investigation of the phase transitions that could sustain hybrid stars is also performed.} 
\maketitle

\section{Introduction}

The analysis of the QCD phase diagram points to a deconfined quark phase standing from the region of high temperatures and low densities down to the region of low temperatures and high densities. While lattice QCD (LQCD) can only describe a small part of the diagram with high temperatures and chemical potentials close to zero, effective models have been extensively used to investigate all other regions
\cite{ratti}. 
From the LQCD perspective, the transition between hadronic matter and deconfined quark matter is a crossover. On the other hand, effective models foresee a first order phase transition.
These two lines can only join if a critical end point (CEP), which should be a unique second order transition point, exists in between them.
At low temperatures, the possibility that a quarkyonic phase \cite{quarkyonic} exists, is not overruled. This phase would consist of matter with the chiral symmetry restored or partially restored but still confined. 

But, what if matter is subject to strong magnetic fields, as in heavy ion collisions, for instance? What do we know about the QCD phase diagram? In \cite{PRD85}, it is shown that the critical chemical potential oscillates around the zero magnetic field value for magnetic fields within $10^{17}$ to $10^{18}$ G range. It is also shown that the CEP position is affected, a result corroborated in \cite{marcio}. In both studies, a unique model, the Nambu-Jona-Lasinio, with and without the Polyakov loop, was used to display the transition line between the hadronic and the quark phase.

Although many works have already investigated the hadron-quark phase transition at zero temperature \cite{muller, diToro, cavagnoli,marquez17, Kauan_2019} with two different models, there are no works investigating the possible transition if matter is subject to strong magnetic fields. This is an interesting subject because of the existence of magnetars \cite{Duncan, Duncan2, Duncan3, Usov, Olausen}, which manifest themselves in quite different ways as compared to the traditional pulsars. 
Could these objects become magnetised quark stars? This is the question we try to answer in the present work. We use two different models, a relativistic hadronic model within a mean field approximation (RMF model) to describe hadronic matter \cite{gm1} and a density dependent model \cite{Xia2014} to describe quark matter. 
The motivation behind the use of a density dependent quark model is two-fold: 1) such models were never utilised in previous works regarding the hadron-quark phase transition and 2) the description of quark stars could result in surface densities lower than the one of regular nuclear matter \cite{Xia2014}, which is a clear signal that the phase transition may occur. This suggests that the cases with low surface densities might be more appropriate to the description of hybrid stars instead of quark stars. The study of hybrid stars gained more interest since a model-independent analysis based on the sound velocity in both hadronic and quark matter suggested that the cores of massive neutron stars should be composed of quark matter \cite{eemeli2020}. In order to fully investigate this case, a proper study of the hadron-quark phase transition is in order, further motivating the use of the density dependent model here considered. In this paper, we
restrict ourselves to the zero temperature regime and 
follow the prescription given in \cite{bombacinuc, marquez17}, which assumes that flavor is conserved during the phase transition, but chemical equilibrium is not. 
We first present the main aspects of the formalism used and then show the results for both the effects of strong magnetic fields on the phase transition and for hybrid stars in chemical equilibrium, followed by a discussion.

\section{Formalism}

To consider the effects of an external magnetic field $B$ on fermions, we modify the calculation of thermodynamic quantities of each particle species with non-zero electric charge $q$, at zero temperature, according to the rule
\begin{equation}
\int d^3k \to \frac{|q| B}{(2\pi)^2} \sum_{\nu} \int dk_z ,
\end{equation}
where $k$ is the momentum, $z$ is the direction of the magnetic field and the sum in $\nu$, the discretised orbital angular momentum that the charged particle acquires in the plane transverse to $B$, goes until a maximum (integer) corresponding Landau level for which its momentum is still real, i.e.,
\begin{equation}
\nu \le \nu_{\rm max} = \left\lfloor \frac{\bar E^2 - \bar m^2}{2|q|B} \right\rfloor ,
\end{equation}
where the relation $\nu=n +\frac{1}{2}-\frac{s}{2}\frac{q}{|q|}$ depends on spin and electric charge. The number density of each charged particle is given by
\begin{equation}
    n_i=\sum_{\nu}\frac{\gamma_\nu}{2\pi^2}|q_i|B\bar{k}_{F_i}
\end{equation}
where $\gamma_\nu$ is the degeneracy of each particle taking into account spin and/or number of colors ($\gamma_\nu=2$ for the spin-1/2 baryons and leptons, and $\gamma_\nu=6$ for quarks; note that $\gamma_{\nu=0}$, the degeneracy of the zeroth Landau level, is always half of the usual value for spin 1/2 particles), and $\bar{k}_{F_i}=\sqrt{\mu_i^2-\bar m_i^2}$ is the Fermi momentum of each particle whose mass was modified by the magnetic field from the bare value $m_i$ to $\bar m_i=\sqrt{m_i^2+2|q_i|B\nu}$. The total effective energy of a charged particle with effective mass $\bar m_i$ in this case becomes
\begin{equation}
\bar E_i=\sqrt{k_z^2+\bar m_i^2+2\nu|q|B} .
\end{equation}
 A complete list of thermodynamic quantities for magnetized fermions at both finite and zero temperature is given in Ref.~\cite{Strickland:2012vu}. For non-charged particle, the pressure and energy density expressions take the usual form.

{The main goal of the present work is to observe how the phase transition point changes with the inclusion of magnetic field, specifically when considering a density dependent quark model. For this purpose, effects such as the inclusion of anomalous magnetic moments were neglected, since they would not substantially change the qualitative results. Similarly, we also restrain our study to isotropic matter and magnetisation is not taken into account.}

\subsection{Hadronic Matter Model}

The hadronic matter is described in this work by a  version of the $\sigma-\omega$ hadrodynamics relativistic mean-field (RMF) model, where the strong interaction between hadrons is emulated by the exchange of virtual mesons. In the Lagrangian density level, considering $b$ species of baryons, the RMF model under an electromagnetic field is given by 
\begin{align}
  \mathcal{L}_{\text{RMF}} ={}& \sum_{b} \bar{\psi }_{b}\Big [ \gamma _{\mu }\Big ( D ^{\mu }-g_{\omega b}\omega _{\mu }-\frac{1}{2}g_{\rho b}\vec{\tau }\cdot \vec{\rho }_{\mu } \Big )\nonumber\\&{}-\left ( m_{b}-g_{\sigma b}\sigma \right )\Big ]\psi _{b}  +\frac{1}{2}\left (\partial ^{\mu } \sigma  \partial _{\mu }\sigma - m_{\sigma }^{2}\sigma ^{2} \right ) \nonumber \\
&- \frac{\lambda }{3}\sigma  ^{3}-\frac{\kappa}{4}\sigma  ^{4} -\frac{1}{4}\Omega ^{\mu \nu }\Omega _{\mu \nu } +\frac{1}{2}m_{\omega }^{2}\omega _{\mu }\omega ^{\mu }  \nonumber \\
&-\frac{1}{4}\vec{P}^{\mu \nu }\cdot \vec{P}_{\mu \nu }+\frac{1}{2}m_{\rho }^{2}\vec{\rho }_{\mu }\cdot \vec{\rho }\,^{\mu }-\frac{1}{4}F^{\mu \nu }F_{\mu \nu },\label{nlwm}  
\end{align}
where $D^\mu=i\partial ^{\mu }+q_bA^\mu$ is the covariant derivative, the meson mass is denoted by $m_i$, with $i=\sigma,\omega,\rho$ (respectively, scalar-isoscalar, vector-isoscalar and vector-isovector) and $g_{ib}$ stands for the coupling constant of the interaction of the $i$ meson field with the baryonic field $\psi_b$. The mesonic and  electromagnetic field  strength  tensors are  given  by  their  usual  expressions, see Ref. \cite{maglag}.
In this work, we consider the GM1 version of the model \cite{gm1}, which was fitted to reproduce nuclear saturation properties
employing  scalar self-meson interactions $\sigma^3$ and $\sigma^4$. 
This parameter set does not fulfill all the constraints for nuclear saturation and stellar properties (see Refs. \cite{nlwm,nlwmstellar}), but the fact that the GM1 model has been widely employed in the literature in works that investigate magnetic field effects and can help us with comparisons and guidance (see, e.g., Refs. \cite{EPJC_2020, jcap_2015}) justifies its choice here. 
{The main inadequacy of the GM1 parameterization was considered its prediction of the symmetry energy slope. However, 
the constraint of the symmetry energy slope recently obtained from the neutron skin thickness observations \cite{prex} shows that it is still suitable. Thus,
despite having been published thirty years ago, this parameterization still holds a good prediction power.}
Moreover, the results that will be obtained next have only qualitative meaning and we do not believe that a different parameterization would modify the main conclusions significantly.

Considering the hadronic matter containing the baryon octet (nucleons $N=\{p$, $n\}$ and hyperons $H=\{\Lambda$, $\Sigma^+$, $\Sigma^0$, $\Sigma^-$, $\Xi^0, \Xi^-\}$), the interaction coupling constants between each meson $i$ and baryon type $b$ can be defined as a fraction of the meson-nucleon coupling $g_{iN}$. The particle relative abundances are very sensitive to the hyperon coupling schemes, so, in this work, we adopt the phenomenologically adjusted parameters proposed in Ref. \cite{gm1}.

In order to describe neutron-star matter, we must observe charge neutrality and chemical equilibrium conditions. To reach these constraints, a non-interacting lepton gas, described by the Lagrangian 
\begin{equation}
      \mathcal{L}_{\rm lepton} = \sum_{l} \bar{\psi }_{l}\left ( \gamma _{\mu } D ^{\mu }-m_{l}\right )\psi _{l}
\end{equation}
is included in the description, with $l=e,\mu$.
The equations of motion for each of the fields are obtained from the Lagrangian (\ref{nlwm}) via the usual Euler-Lagrange formalism, employing a mean-field approximation which allows to obtain the energy-momentum tensor for this model. The equation of state (EoS) is obtained from the energy-momentum tensor once the field equations are solved, such that the total energy density $E$ is given by
\begin{equation}
    E=E_m + \frac{B^2}{2},
\end{equation}    
    where
\begin{equation}
    E_m=E_{q=0}+E_{q\ne0}+E_{\rm meson}+E_{\rm lepton}, \label{endens_rmf}
\end{equation}    
and $E_{q=0}$ and $E_{q\ne0}$ are the energy densities of neutral and charged baryons, respectively, and are given by
\begin{equation}
E_{q=0} = \sum_{b|q=0}\frac{1}{\pi^2}\int_0^{k_{F_b}}dk\,k^2\sqrt{k^2+M_b^2},
\end{equation}
where $M_b=m_{b}-g_{\sigma b}\sigma $, and
\begin{equation}
E_{q\ne0}=\sum_{b|q\ne0}\frac{|q| B}{(2\pi)^2} \sum_{\nu} \int_0^{k_{z\,F_b}}\!\! dk_z\sqrt{k_z^2+\bar{M}_b^2},
\end{equation}
where $\bar M_b=\bar m_{b}-g_{\sigma b}\sigma $, and
\begin{equation}
    E_{\rm meson}=\frac{1}{2}m_{\sigma }^{2}\sigma ^{2}   -\frac{1}{2}m_{\omega }^{2}\omega ^{2 } -\frac{1}{2} m_{\rho }^{2}\rho^{2 }+ \frac{\lambda}{3}\sigma  ^{3}+\frac{\kappa}{4}^{4},
\end{equation}
is the mesonic fields contribution, $E_{\rm lepton}$ is the energy density of a free and magnetized fermion gas
\cite{Strickland:2012vu}
, and the last term
accounts for the electromagnetic field contribution (here, in Heaviside-Lorentz natural units).
The pressure is given by the general thermodynamic expression 
\begin{equation}
    p_m = -E_m+ \sum_i \mu_i n_i \label{pndens_rmf}
\end{equation}
with the index $i$ accounting for the baryons and leptons. In the present work we assume that the system is isotropic. For a discussion on
different formalisms dealing with the inclusion of strong magnetic fields (anisotropy, chaotic field, LORENE code), the interested reader can look at \cite{jcap_2015, smf} and references therein.

In the following calculations, the particle fractions $Y_i=n_i/n_B$, where $n_B$ is the baryonic density, 
are determined from the neutron and electron chemical potentials
\begin{equation}
    \mu_b=\mu_n-q_b\mu_e, \label{betaeq}
\end{equation}
where $q_b$ is the electric charge of the baryon $b$, and $\mu_\mu=\mu_e$.

When describing neutron star matter under strong magnetic fields, it is not unusual in the literature to consider $B$ varying with density according to some scaling relation (see, e.g., Refs. \cite{Duncan, Chakrabarty}), to make the magnetic field range from $10^{18} ~\rm G$ in the center of the star to $10^{15} ~\rm G$  on its surface. 
In the present work, the magnetic field is taken as constant to avoid the violation of one of Maxwell's equations, as shown in \cite{maxwellviol}.

\subsection{Quark Matter Model}

In this work, the density dependent quark mass (DDQM) model is utilized to describe quark matter. Under this approach, the quark confinement is described by the density dependence introduced in the quark masses:

\begin{equation}
    m_i = m_{i0} + \frac{D}{n_b^{1/3}} + Cn_b^{1/3} = m_{i0} + m_I,
    \label{masses}
\end{equation}
where $m_{i0}$ (i = u, d, s) is the current mass of the $i$th quark, $n_b$ is the baryon number density and $m_I$ is the density dependent term that mimicks the interaction between quarks. The model has two free adjustable parameters: $D$, that dictates linear confinement; and $C$, that is responsible for leading-order perturbative interactions \cite{Xia2014}.

Whenever a density dependent term is introduced, the issue of thermodynamic consistency arises. To overcome this problem, we follow the formalism of \cite{Xia2014}, that presents a thermodynamically consistent DDQM model. The introduction of magnetic field is done in a similar way to \cite{Isayev2015}, where the density dependent MIT Bag Model was thermodynamically treated. Under this approach, magnetised quark matter will be treated as being uniform and permeated by an external uniform magnetic field. 

At zero temperature, the differential fundamental relation holds

\begin{equation}
    \text{d}E_m = \sum_i \mu_i\text{d}n_i,
    \label{diff-fundamental-eq}
\end{equation}
where $E_m$ is the matter contribution to the energy density of the system, $\mu_i$ are the particles chemical potentials and $n_i$ are the particle densities. 

One way of overcoming thermodynamic inconsistency is by the introduction of effective chemical potentials \cite{effective-chem-pot}. Under this perspective, the energy density can be viewed as the one of a free system with particle masses $m_i(n_b)$  and effective chemical potentials $\mu_i^*$

\begin{equation}
    E_m = \Omega_m^0 (\{\mu_i^*\},m_i,B) + \sum_i \mu_i^* n_i,
    \label{free-system-fundamental-eq}
\end{equation}
where $\Omega_m^0$ is the thermodynamic potential of a free system in the presence of an external magnetic field. At a fixed $B$, the differential form of Eq. \eqref{free-system-fundamental-eq} is

\begin{equation}
    \text{d}E_m = \text{d}\Omega_m^0 + \sum_i \mu_i^* \text{d} n_i + \sum_i n_i \text{d}\mu_i^*.
    \label{initial-diff-free-system}
\end{equation}
Explicitly, we can write d$\Omega_m^0$ as

\begin{equation}
    \text{d} \Omega_m^0 = \sum_i \frac{\partial \Omega_m^0}{\partial \mu_i^*} \text{d}\mu_i^* + \sum_i \frac{\partial \Omega_m^0}{\partial m_i}\text{d}m_i
\end{equation}
with

\begin{equation}
    \text{d}m_i = \sum_j \frac{\partial m_i}{\partial n_j} \text{d}n_j,
\end{equation}
where the densities are connected to the effective chemical potentials by
\begin{equation}
    n_i = -\frac{\partial \Omega_m^0}{\partial \mu_i^*}
\end{equation}
to ensure thermodynamic consistency. 

Eq. \eqref{initial-diff-free-system} can then be rewritten as

\begin{equation}
    \text{d}E_m = \sum_i \left(\mu_i^* + \sum_j \frac{\partial \Omega_m^0}{\partial m_j} \frac{\partial m_j}{\partial n_i}\right) \text{d}n_i,
    \label{diff-free-system}
\end{equation}
that should be consistent with the fundamental equation. Comparing eqs. \eqref{diff-fundamental-eq} and \eqref{diff-free-system}, one gets the relation between the real and the effective chemical potentials

\begin{equation}
    \mu_i = \mu_i^* + \sum_j \frac{\partial \Omega_m^0}{\partial m_j} \frac{\partial m_j}{\partial n_i}.
\end{equation}

Considering magnetized quark matter to be an isotropic gas, the matter contribution to the pressure, $p_m$, is then given by
\begin{equation}
    p_m = -E_m + \sum_i \mu_i n_i. 
\end{equation}
The introduction of the effective chemical potentials through Eq. \eqref{free-system-fundamental-eq} gives
\begin{align}
    p_m ={}& -\Omega_m^0 + \sum_i (\mu_i - \mu_i^*)n_i\nonumber\\
    ={}& -\Omega_m^0 + \sum_{i,j} \frac{\partial \Omega_m^0}{\partial m_j}n_i\frac{\partial m_j}{\partial n_i}.
    \label{pressure-quarks}
\end{align}

One can note that, from basic thermodynamics, we can write the matter contribution to the thermodynamic potential, $\Omega_m$, as

\begin{equation}
    \Omega_m = E_m - \sum_i \mu_i n_i,
\end{equation}
and plugging Eq. \eqref{free-system-fundamental-eq} yields

\begin{equation}
    \Omega_m = \Omega_m^0 - \sum_i(\mu_i - \mu_i^*)n_i,
\end{equation}
so that the thermodynamic relation $\Omega_m = - p_m$ still holds.

The EoS of the system can then be obtained by taking into account the electromagnetic field contribution. In Heaviside-Lorentz natural units, the total energy density $E$ and the pressure $p$ are given by

\begin{equation}
    E = E_m + \frac{B^2}{2} = \Omega_m^0 +\sum_i \mu_i^*n_i + \frac{B^2}{2}
\end{equation}

and

\begin{equation}
    p = p_m + \frac{B^2}{2} = -\Omega_m^0 + \sum_{i,j} \frac{\partial \Omega_m^0}{\partial m_j}n_i \frac{\partial m_j}{\partial n_i} + \frac{B^2}{2}.
\end{equation}

It is worth noting that there are several different mass scaling relations for DDQM models, such as the inverse linear scaling \cite{linear-mass-scaling} and the simple cubic scaling \cite{single-cubic-scaling,single-cubic-scaling2}. All of them remain thermodynamically consistent within this approach, and the main advantage of using the scaling relation of Eq. \eqref{masses} is that the inclusion of the parameter $C$ enables the attainment of more massive stars \cite{Backes2020}.

\begin{figure}[!h]
\begin{center}
\includegraphics[angle=270,width=8cm] {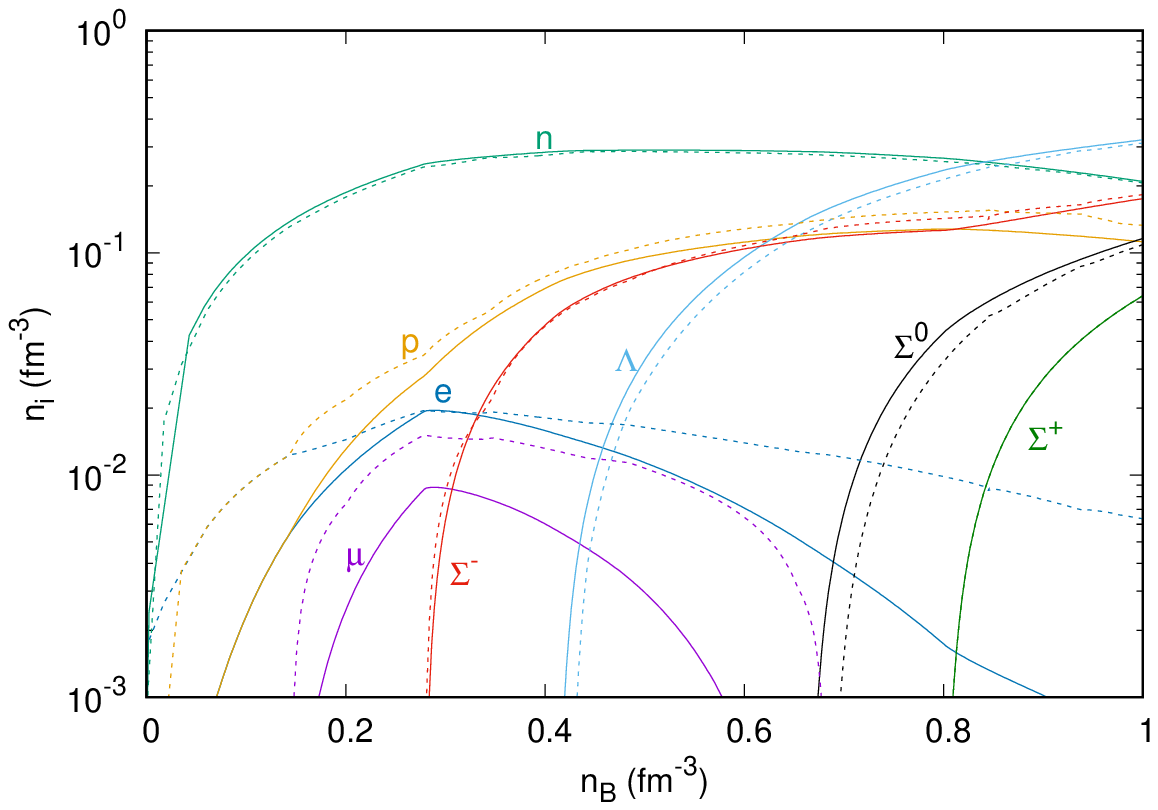}
\includegraphics[angle=270,width=8cm] {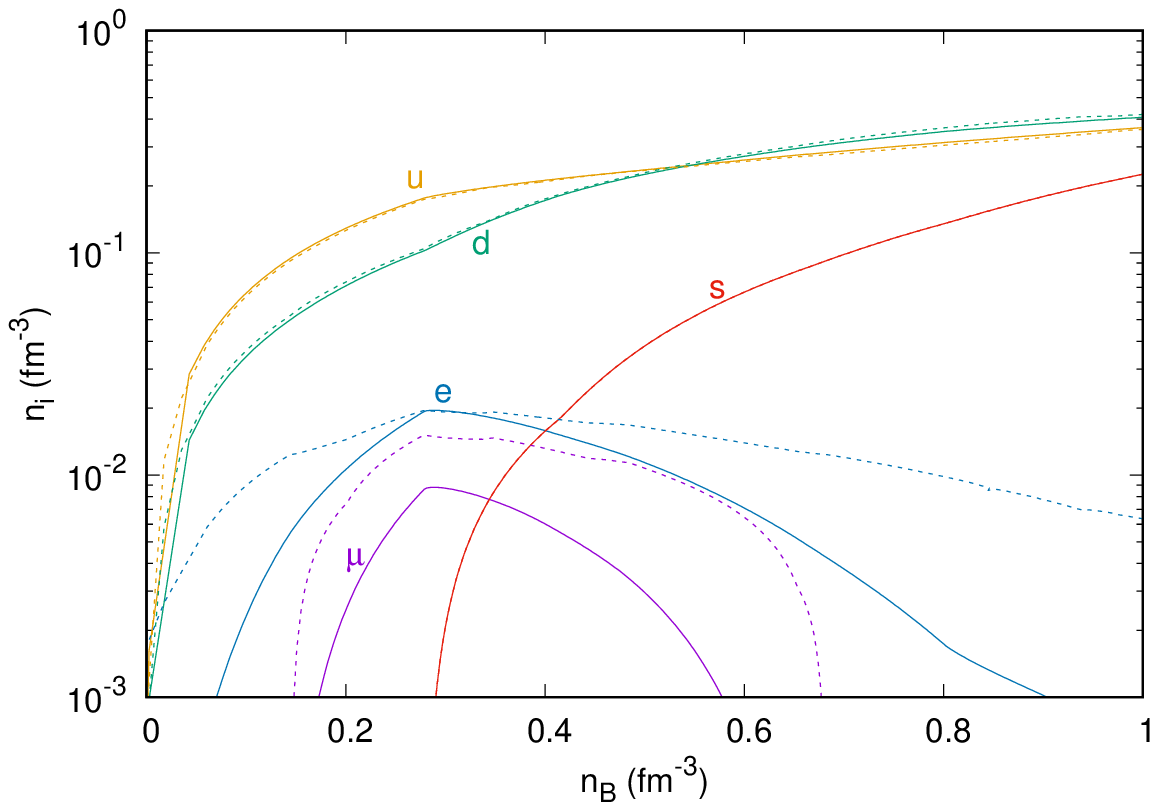}
\caption{Baryon species populations from  the RMF model for GM1 parameterization (above) and the respective quark fractions (below). Full lines show results without magnetic fields, while dashed lines show results including a magnetic field of $B=3\times10^{18}$ G.} \label{populations}
\end{center}
\end{figure}

\section{Results and discussions}

\begin{figure*}[!ht]
\begin{center}
\includegraphics[angle=270,width=8.9cm] {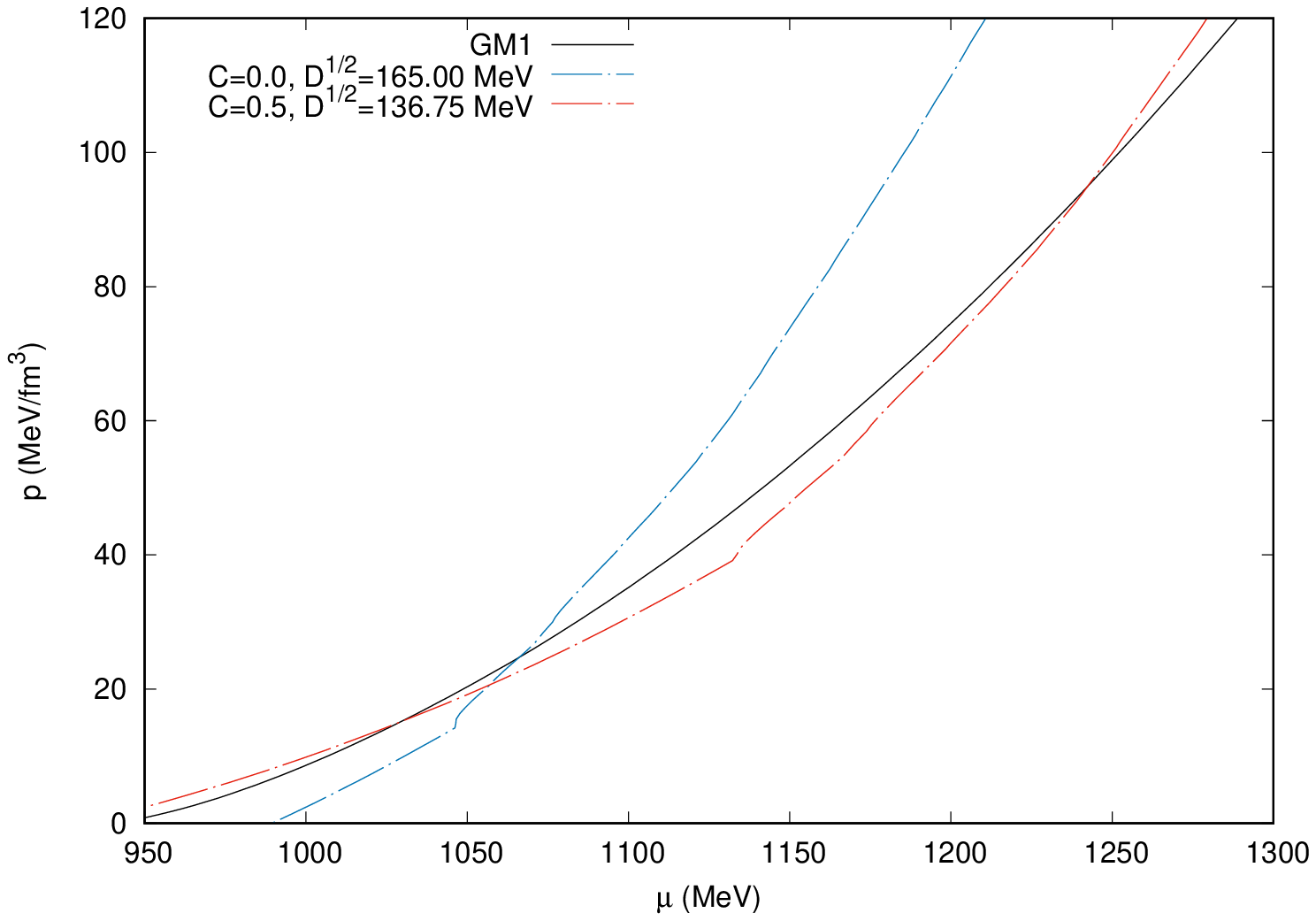}
\includegraphics[angle=270,width=8.9cm] {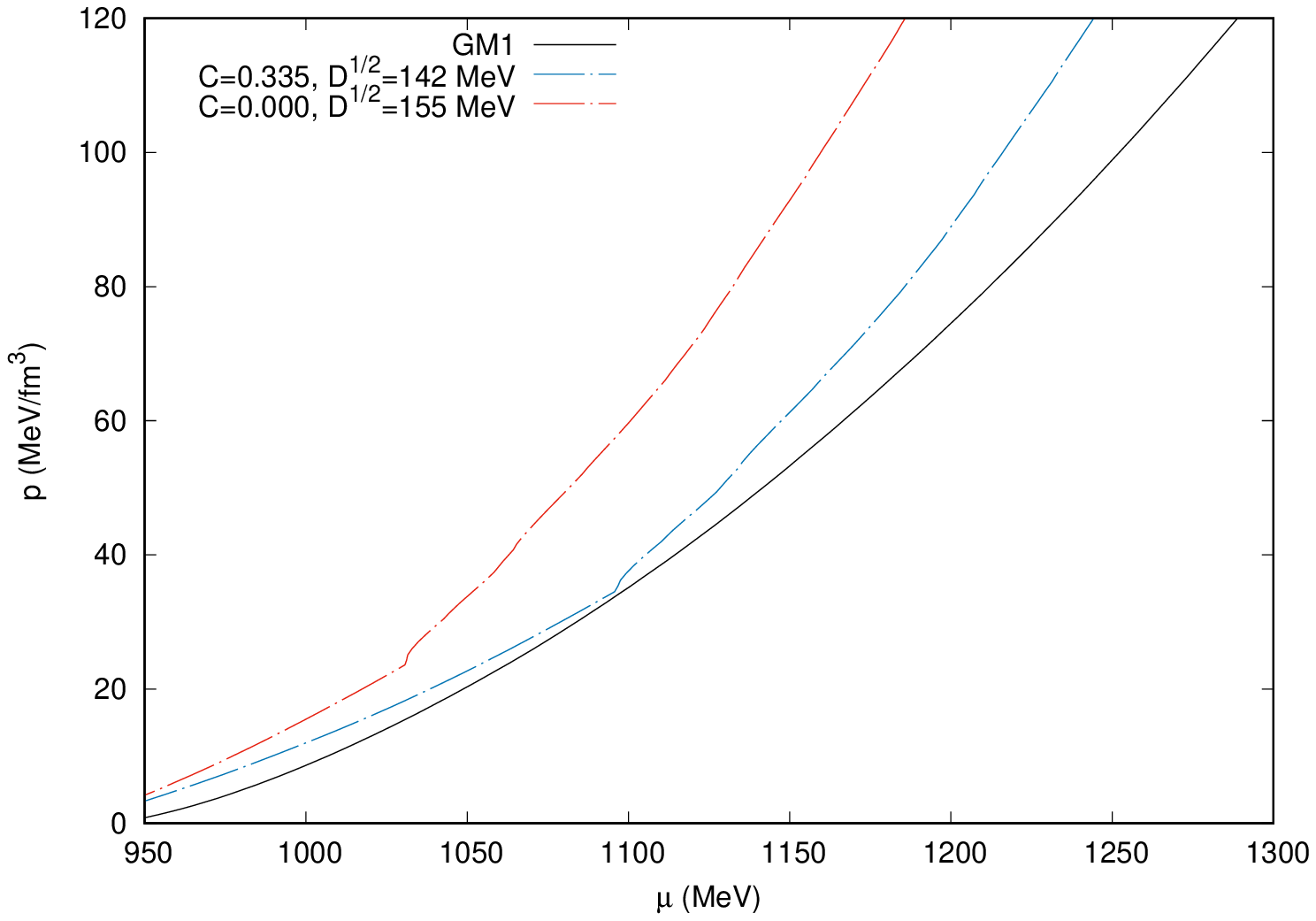}
\caption{Example of parameter sets that allow (left panel) and do not allow (right panel) the hadron-quark phase transition to occur at $B$ = $3 \times 10^{18}$ G.} 
\label{crossing}
\end{center}
\end{figure*}

According to the Bodmer-Witten conjecture \cite{bodmer,witten}, under certain circumstances, the electrically neutral and in chemical equilibrium  hadronic matter is metastable and can be converted into an energetically favored, deconfined quark phase.
The deconfinement of the hadronic matter into the quark phase must occur in the strong interaction time scale, which is many orders of magnitude smaller than the weak interaction time scale \cite{nucleation,critmass}. It implies that the relative flavor fractions must be conserved during the phase transition,
\begin{equation}
 Y_q=\frac{1}{3}\sum_i n_{qi} Y_i , \label{eq:flavour}
\end{equation}
where the baryonic number densities of each particle species $ n_i=Y_in_B$ are related by the number $n_{qi}$ of $q$ flavored quark constituents of baryon $ i $ \cite {nucleation,marquez17}.  As we consider that the total baryonic mass and the lepton number are also conserved, Eq. \eqref{eq:flavour} determines the composition of the resulting quark phase from the initial hadronic matter in chemical equilibrium. 
Quark matter will not be in $ \beta $-equilibrium, but the process preserves charge neutrality, as explained in Ref. \cite{marquez17}.
Figure \ref{populations} shows the particle population obtained from the  RMF model, considering the equilibrium conditions, and the respective quark matter fractions associated with this hadronic distribution. Notice that the leptons are present in both configurations and are more affected by the magnetic field than the baryons and the quarks. In fact, the lepton contribution is defined in the hadronic phase, as stated above.

The deconfinement transition is assumed to be a first-order phase transition. The thermodynamic description of this kind of process can be obtained from the matching of the equations of state for the two phases. The transition can happen after the over-pressured metastable matter reaches the phase coexistence point, defined according to the Gibbs criteria as, 
\begin{equation}
\begin{gathered}
  p^{(i)}=p^{(f)}=p_0,\\
  \mu^{(i)}(p_0)=\mu^{(f)}(p_0)=\mu_0,
\end{gathered} \label{eq:gibbscon}
\end{equation}
for the transition between the initial ($i$) and final ($f$) phases considered homogeneous, with
\begin{equation}
 \mu^{ (i,f)}=\frac{\varepsilon^{(i,f)}+P^{(i,f)}}{n_B^{(i,f)}},
\end{equation}
where $\varepsilon^{(i,f)}$, $p^{(i,f)}$ and $n_B^{(i,f)}$ are the total energy density, pressure and number density, obtained from the effective model EoS. \cite{critmass}.
These conditions leave the values of $p_0$ and $\mu_0$ to be determined from the equations of state of both phases. Notice that in the results that follow, we neglect the $B^2/2$ term in the pressure of both models and verify only the crossing of the curves related to hadronic and quark matter, otherwise, it would be impossible to compare our results with the ones obtained with non-magnetised matter, since the contribution from the pure magnetic field (for a fixed value of $B$) is very large as compared with the contribution of magnetised matter. {This way, the equations utilised for computing the EoS of hadronic matter are given by \eqref{endens_rmf} and \eqref{pndens_rmf}, and for quark matter \eqref{free-system-fundamental-eq} and \eqref{pressure-quarks}.}

Since we utilise only one parameterization to describe the hadronic phase, the condition of coexistence of phases may or may not be satisfied depending only on the DDQM model free parameters, $C$ and $D$. The procedure for checking the Gibbs criteria is graphically shown in Figure \ref{crossing}, for sets of parameters that allow or do not allow the phase transition to occur. For some sets of parameters, such as $C$ = 0.5 and $\sqrt{D}$ = 136.75 MeV, the conditions of Eq. \eqref{eq:gibbscon} are satisfied more than once. Whenever such double crossing occurs, only the hadron-quark phase transition is considered, since a quark-hadron phase transition is not expected to exist.

In regard to the DDQM models free parameters, the adequate stability window must be taken into account. We restrict our study to the sets of parameters that satisfy the Bodmer-Witten conjecture or are barely outside of the stability window of non magnetised strange quark matter \cite{Backes2020}, which is reasonable since the binding energy of magnetised matter is lower than the one of non magnetised matter \cite{veronica,gonzalez2009,wen2012}.

\begin{table}[!h]
    \centering
    \begin{tabular}{c|c|c|c|}
       & B = 0 & B = $3 \times 10^{18}$ G & B-W \\ \hline
$C$ = 0  & no crossing & no crossing & yes  \\ 
$\sqrt{D}$ = 155 MeV &  &  & \\ \hline
$C$ = 0  & $\mu_0$ = 960 & $\mu_0$ = 958 & yes  \\ 
$\sqrt{D}$ = 158.5 MeV & $p_0$ = 1.55 & $p_0$ = 1.80 & \\ \hline
$C$ = 0  & $\mu_0$ = 1062 & $\mu_0$ = 1066 & no  \\ 
$\sqrt{D}$ = 165 MeV & $p_0$ = 21.98 & $p_0$ = 24.70 & \\ \hline
$C$ = 0.23  & $\mu_0$ = 1130 & $\mu_0$ = 1145 & no  \\ 
$\sqrt{D}$ = 155 MeV & $p_0$ = 43.62 & $p_0$ = 51.32 & \\ \hline
$C$ = 0.365  & $\mu_0$ = 1105 & $\mu_0$ = 1109 & yes  \\ 
$\sqrt{D}$ = 142 MeV & $p_0$ = 34.98 & $p_0$ = 38.30 & \\ \hline
$C$ = 0.5  & $\mu_0$ = 1202 & $\mu_0$ = 1242 & yes \\ 
$\sqrt{D}$ = 135.75 MeV & $p_0$ = 72.66 & $p_0$ = 94.93 & \\ \hline
$C$ = 0.68  & $\mu_0$ = 1440 & $\mu_0$ = 1475 & yes \\ 
$\sqrt{D}$ = 130 MeV & $p_0$ = 215.50 & $p_0$ = 247.53 &  \\ \hline
    \end{tabular}
    
    \caption{Values for $\mu_0$ (in MeV) and $p_0$ (in MeV/fm$^{3}$) for which the conditions of phase coexistence are satisfied at T = 0. Results are shown for sets of parameters $C$ and $D$ within and outside of the stability window of SQM, for both magnetised and demagnetised matter. The latter column specifies whether or not the Bodmer-Witten conjecture is satisfied.
    }
    \label{tab-phase-coexistence}
\end{table}

The results for $\mu_0$ and $p_0$ are summarized in Table \ref{tab-phase-coexistence}, where it can be seen that the inclusion of magnetic field shifts the coexistence point towards higher pressures and generally, also towards higher chemical potentials. The exception (second line on the table) may be due to numerical uncertainties implicit to these simple models.
This shift tends to be higher when the perturbative parameter becomes larger. This result goes in line with the general notion that the EoS stiffens as the magnetic field increases. It was also observed that for a fixed value of $D$, the coexistence point occurs at higher pressures when $C$ increases. Similarly, for a fixed value of $C$, the coexistence point occurs at higher pressures when $D$ increases.

In Ref. \cite{Xia2014}, the same DDQM model used in this work is applied to the study of strange stars. It was shown that the surface density of stars described with high $C$ parameters is even lower than nuclear saturation density, which points to the existence of a phase transition. By analysing the results of Table \ref{tab-phase-coexistence}, it is noticeable that as $C$ increases, the coexistence point indeed occurs at higher pressures. This result corroborates the previous findings, suggesting that whenever a large perturbative parameter is considered, the DDQM model could be more suitable for the description of hybrid stars instead of strange ones. {Since the issue of hadron-quark phase transition with the DDQM model has already been  addressed,} we further analyse this possibility. 

{For large values of $C$,  a double crossing can occur when one investigates the Gibbs criteria of phase coexistence, as shown in Figure \ref{crossing}. 
Whenever there is a double crossing, the first coexistence points (that predict a quark-hadron phase transition) are always at low enough densities, below the cusp that can be observed in the EoS at the point where strange quarks first appear, so that there are no strange quarks. 
Thus, not only such a transition is not expected to exist from a phenomenological point of view, but it is also not favorable since two flavor quark matter will always be unstable against nuclear matter, respecting the Bodmer-Witten hypothesis \cite{bodmer,witten} that was already considered for such quark model parameters \cite{Backes2020}. Therefore, the first crossing point must be disregarded so that matter is confined at the low density regime.}

\begin{figure}[!ht]
\begin{center}
\includegraphics[angle=270,width=8cm] {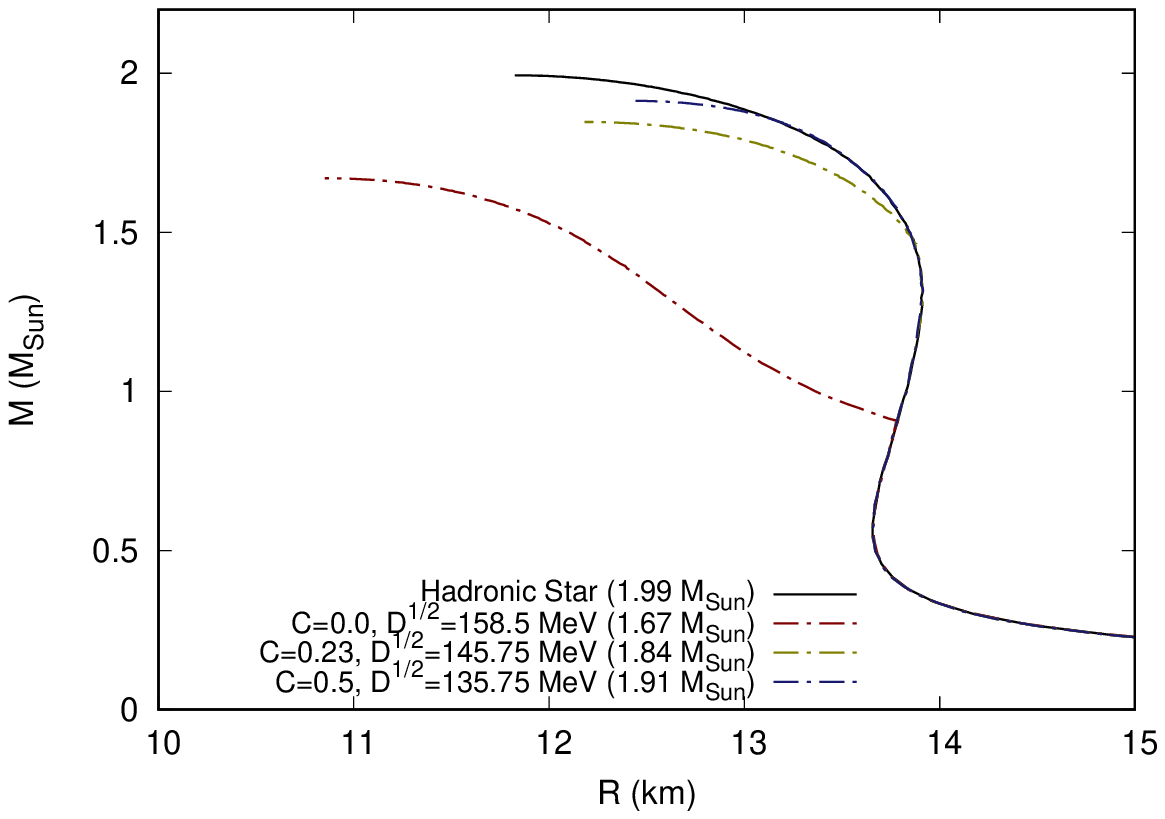}
\caption{Mass-radius diagram for hybrid EoS with chemical equilibrium in both phases, showing results without magnetic field effects. The maximum stellar masses are indicated for all cases.} \label{tov}
\end{center}
\end{figure}

An EoS that describes hybrid stars can be built by a Maxwell construction, interpolating the hadronic and the quark EoS at the coexistence point shown in Table \ref{tab-phase-coexistence}. 
Figure \ref{tov} shows mass-radius curves produced from inserting some EoS obtained with a Maxwell construction into the Tolman–Oppenheimer–Volkoff (TOV) equations \cite{TOVt,TOVov} and adding the BPS EoS \cite{bps} for the low-density region of the crust. Our results here consider only stellar matter without magnetic field effects, since the use of the TOV-like equations for magnetised matter requires the solution of a more complicated system of equations in general relativity \cite{Bocquet:1995je,Cardall:2000bs,Frieben:2012dz,Pili:2014npa}, which goes beyond the objectives of the present work. Pure strange stars would show maximum masses close to the ones of the respective hybrid star (the values for strange star are $M=1.63 ~\rm M_{Sun}$ for the DDQM model parameters $C=0.0$, $D^{1/2}=158.5$ MeV; $M=1.78 ~\rm M_{Sun}$ for $C=0.23$, $D^{1/2}=145.75$ MeV; and $M=1.92 ~\rm M_{Sun}$ for $C=0.5$, $D^{1/2}=135.75$ MeV), but the profile of the mass-radius curve would be different, as no crust is expected to remain in quark stars \cite{Melrose, Melrose2}{, although such possibility can also be considered \cite{haensel}}.

Following the stellar evolution scenario proposed by Refs. \cite{marquez17,Kauan_2019}, we can assume the compact star as being initially a pure hadronic metastable star in the early stages after its emergence. In this stage, the equilibrium conditions are reached through the first deleptonization and cooling, and the resulting objects are the ones described by the black curve in Fig. \ref{tov}. After a finite time interval,
this metastable configuration can decay into an energetically more favorable one, and, according to the Bodmer-Witten conjecture, it can be reached by the quark deconfinement. So, the conversion of a metastable hadronic star into a hybrid (or strange) star can take place via a first order transition. The transition dynamics we consider in this work assumes flavor conservation in the first moment, as imposed by Eq. \eqref{eq:flavour}, 
and this condition is taken to determine the phase coexistence points. Finally, the quark matter would seek the chemical equilibrium soon after its formation, because this is the stable configuration of stellar matter, and it justifies the use of this configuration in Fig. \ref{tov}.

Notice that a deconfined quark core starts appearing at a very low masses ($M=0.90 ~\rm M_{Sun}$) when the parameters $C=0.0$, $D^{1/2}=158.5$ MeV are considered. On the other hand, the transition does not take place in the star matter density threshold for the parameters $C=0.68$, $D^{1/2}=130$ MeV. 
{In the stellar evolution scenario presented above, the conversion of a metastable hadronic star to a strange or hybrid star could take place if the initial object sustains a central pressure larger than the coexistence pressure of the hadron and quark phases ($p_0$).
Yet, when the metastable hadronic star matter is overpressured enough to allow the appearance of a quark core, there must exist a stable final compact object whose central density is that of the EoS latter phase, i.e., we must also have a final object whose central pressure (which is the higher pressure present in this object) is higher or equal $p_0$, or the metastable star will not have a stable compact star configuration to decay (if not a black hole).}  In the case of parameters $C=0.68$, $D^{1/2}=130$ MeV, this constraint is not fulfilled, and the maximum stable star  would be a pure hadronic star  of $M=1.96 ~\rm M_{Sun}$ and $R=12.42$ km, never sustaining a quark core beyond this point, 
following the trend of hybrid star maximum masses shown in Fig. \ref{tov}. 
This set fulfills the thermodynamic criteria for the phase transition, but the astrophysical conditions do not allow the formation of a stable hybrid star.

\section{Conclusions}

In the present work we have analysed the conditions for a phase transition from a hadronic to a quark phase with the help of two different models at zero temperature and under the influence of a very strong magnetic field, possibly present in the core of magnetars.
With the chosen models (GM1 parameterization with hyperon-meson couplings fixed phenomenologically for the hadronic matter and DDQM for the quark phase) we
have, in general, obtained an increase of pressure and chemical potential as compared with the transition point of non-magnetised matter. 

For the sake of completeness, 
whenever the phase transition is possible, we have checked whether stable hybrid stars would be allowed and if so, if their cores could indeed contain deconfined quarks. This part of the work was performed considering non-magnetised matter. An investigation with an appropriate code written for magnetised EoS (LORENE) remains to be done, but the results presented here can already give us a hint of the physical scenario as far as qualitative aspects are concerned.

After an extensive analysis of the hadron-quark phase transition, it was observed that although some sets of parameters with large $C$ values of the DDQM model are within the stable SQM regime and produce high stellar masses for strange stars \cite{Backes2020}, they are neither suitable for the description of hybrid stars or quark stars if one believes on the conversion process discussed above.

The present work is the seed for the study of the whole magnetised matter QCD phase diagram, which requires the introduction of temperature {and anomalous magnetic moment} in both models, works already in progress.

\section*{Aknowledgements}

 This  work is a part of the project INCT-FNA Proc. No. 464898/2014-5. D.P.M. and K.D.M. are  partially supported by Conselho Nacional de Desenvolvimento Cient\'ifico e  Tecnol\'ogico (CNPq/Brazil) respectively  under grant \\ 301155.2017-8 and with a doctorate scholarship. B.C.B is supported by Coordenação de Aperfeiçoamanto de Pessoal de Nível Superior (CAPES) with a M.Sc. scholarship. B.C.B. thanks fruitful discussions with Eduardo Hafemann about the convergence of the quark matter numerical code.

\bibliographystyle{unsrt}
\bibliography{paper}

\end{document}